\documentclass[12pt]{article}

\usepackage{amssymb,amsthm,amsfonts,psfig,epsfig, multicol,mathrsfs,axodraw}
 
\def\BrightRed  {}
\def\Black{}
\def\Green{} % PANTONE 323
\def\Blue {}
\def\gappeq{\mathrel{\rlap {\raise.5ex\hbox{$>$}} {\lower.5ex\hbox{$\sim$}}}}
\def\lappeq{\mathrel{\rlap{\raise.5ex\hbox{$<$}} {\lower.5ex\hbox{$\sim$}}}}
\def\beq{\begin{equation}} \def\eeq{\end{equation}} 
\def\bea{\begin{eqnarray}} \def\eea{\end{eqnarray}}
\def\bq{\begin{quote}} \def\eq{\end{quote}}
\def\bc{\begin{center}} \def\ec{\end{center}}

\def\nn{\nonumber}

\begin{document} 

%%%%%

\pagestyle{empty}
\def\thefootnote{\fnsymbol{footnote}}

\begin{flushright} ROMA1-TH/1410-05  \end{flushright}
\vskip 2cm

\bc {\Large \bf  A Maximal Atmospheric Mixing \\ \vskip .2cm  from a Maximal CP Violating Phase}  \ec
\vspace*{2cm} 

\centerline{\bf \large Isabella Masina} 
\vskip 1cm

\bc { \em Centro Studi e Ricerche "E. Fermi", Via Panisperna 89/A, Rome, Italy \\ \vskip .1cm  
and \\ \vskip .1cm  INFN, Sezione di Roma, P.le A. Moro 2, Rome, Italy} \ec
\vskip 2cm

\centerline{\bf Abstract} 
\vskip.2cm
We point out an elegant mechanism to predict a maximal atmospheric angle,
which is based on a maximal CP violating phase difference between second and third lepton families 
in the flavour symmetry basis.
In this framework, a discussion of the general formulas for $\theta_{12}$, $|U_{e3}|$, $\delta$ 
and their possible correlations in some limiting cases is provided.       
We also present an explicit realisation in terms of an $SO(3)$ flavour symmetry model. 
%%%%%%

\newpage
\setcounter{page}{1}
\pagestyle{plain}
\def\thefootnote{\arabic{footnote}}
\setcounter{footnote}{0}

%%%%%%%%%%%%%%%  START %%%%%%%%%%%%%%%%%%%%%%%%%%%%%%%%%

\section{Introduction}

The atmospheric mixing angle has experimentally turn out to be quite large \cite{exper} and,
according to recent analysis \cite{Fogli}, its $1\sigma$ range is $\theta_{23}=38^\circ -47^\circ$.
From the theoretical point of view this invites to speculate on the possibility 
%$\theta_{23}=\pi/4$, {\it i.e.} 
of maximal flavour violation for second and third lepton families.
In flavour model building, achieving a maximal $\theta_{23}$ is far from trivial,
taking also into account that $|U_{e3}|$ is small and the solar mixing angle 
large but not maximal: $\theta_{13} \le 7^\circ$ and $\theta_{12}= 33^\circ -36^\circ$ at 1$\sigma$ \cite{Fogli}. 
Our aim is to investigate which are the most natural mechanisms to generate a maximal atmospheric 
mixing\footnote{To be quantitative (and rather subjective),
we are going to ask $|1-\tan\theta_{23}|\lesssim 5\%$, corresponding to an uncertainty of about $2^\circ$
in $\theta_{23}$.}.

If an underlying flavour symmetry exists, it selects a {\it privileged flavour basis} for fermion mass matrices. 
The lepton mixing matrix results from combining the unitary matrices which - in that basis -
diagonalise left-handed charged lepton and neutrino mass matrices respectively, $U_{MNS}=U_e^\dagger U_\nu$.
Obtaining a maximal atmospheric angle because of a conspiracy between many large mixings 
present in $U_e$ and $U_\nu$ appears to be quite a fortuitous explanation,
especially in the case of effective neutrino masses as in the seesaw 
mechanism\footnote{Exceptions are the $A_4$ models \cite{A4}, 
where tunings are eventually displaced in the neutrino spectrum.}.
A better starting point for a flavor model is to predict one between $\theta^e_{23}$ and $\theta^\nu_{23}$ to be maximal, 
so that the goal is reached when the other parameters present in $U_e^\dagger U_\nu$ 
marginally affect this maximal 23-angle.

Models so far proposed \cite{others} along these lines adopted the strategy of having, 
in the flavour symmetry basis, 
a maximal $\theta^{\nu}_{23}$ together with a negligibly small $\theta^{e}_{23}$, or viceversa. 
Since the upper bound on $\theta_{13}$ naturally suggests $\theta^e_{12}$ and $\theta^e_{13}$  
to be small and since the latter affect $\theta_{23}$ at second order,
this is in principle a simple and effective framework to end up with a maximal atmospheric angle.
However, the drawback in model building is the difficulty in managing such a huge hierarchy among 
$\theta^{e}_{23}$ and $\theta^{\nu}_{23}$.

In this letter we point out an alternative mechanism to achieve a maximal atmospheric angle, 
which is based on the presence of a maximal CP violating phase difference between second and third lepton families
and which does not require any particular hierarchy among $\theta^{e}_{23}$ and $\theta^{\nu}_{23}$,
provided that one of them is maximal. 
If $\theta^e_{12}$ and $\theta^e_{13}$ are small
as suggested by the bound on $\theta_{13}$, then $\theta_{23}=\pi/4$ is robustly predicted. This mechanism 
is based on maximal CP violation in the sense that,
denoting by $g/\sqrt{2}~ (e^{-i w_2}  {\bar\mu}_L ~\gamma^\lambda  \nu_\mu + {\bar \tau}_L ~\gamma^\lambda \nu_\tau)~ 
W^-_\lambda +$ h.c. the weak charged currents of the second and third lepton families
in the flavour symmetry basis (before doing the rotations in the $\mu-\tau$ and $\nu_\mu-\nu_\tau$ planes to go 
in the mass eigenstate basis), it requires the phase difference $w_2$ to be $\pm \pi/2$. 
For three families and Majorana neutrinos, $U_{MNS}$ contains three CP violating phases, 
which turn out to be complicated functions of $w_2$ and of other phases potentially present. 
We will focus in particular on the connection between $\delta$ and $w_2$ 
and on the expectations for $|U_{e3}|$ and $\theta_{12}$.
This mechanism also has remarkable analogies with the quark sector.

%%%%%%%%%%%%%%%%%%%%%%%%%%%%%%%%%%%%%%%%%%%%%%%%%%%

\section{On the Origin of the Atmospheric Mixing}

In the basis of the unknown flavour symmetry the leptonic sector is described by
\bea
& &{\cal L}~ = -\frac{1}{2}~\nu^T m_{\nu}^{eff} \nu ~-~ \bar e_R^{T} m_e e_L ~
+~\frac{g}{\sqrt{2}} ~\bar e_L^T \gamma^\lambda \nu ~W^-_\lambda  ~+~ \mathrm{h.c.}~~~~~\label{lagr}~\\
& & ~~~~~~~~~~~m_{\nu}^{eff}= U_\nu^* \hat m_\nu U_\nu^\dagger ~~,~~~~~~ m_e = U^R_e \hat m_e {U^L_e}^\dagger \nn
\eea
where a hat is placed over a diagonal matrix with real positive eigenvalues 
whose order is established conventionally by requiring $|m^2_2 -m^2_3| \ge m^2_2 - m^2_1 \ge 0$,
and the $U$'s are unitary matrices. 
The MNS mixing matrix is $U_{MNS}={U^L_e}^\dagger U_\nu$.
We find it convenient to write unitary matrices in terms of a matrix in the standard CKM parameterization \cite{RPP}
multiplied at right and left by diagonal matrices with five independent phases, defined for definiteness according to
\beq
U_{\ell} = e^{i \alpha_{\ell}} ~ {\cal W}_{\ell}~ U^{(s)}_{\ell}~ {\cal V}_{\ell} ~~~~~~~~\ell=e,\nu, \mathrm{MNS}  ~~~~~
\label{par}
\eeq
where, omitting the $\ell$ index, 
${\cal W}=$ diag$(e^{i (w_1+w_2)}, e^{i w_2},1)$, ~${\cal V}=$ diag$(e^{i (v_1+v_2)}, e^{i v_2},1)$,~
$U^{(s)}=R(\theta_{23}) \Gamma_\delta R(\theta_{13}) \Gamma_\delta^\dagger R(\theta_{12})$,
$\Gamma_\delta =$diag$(1,1,e^{i \delta})$,  
angles belong to the first quadrant and phases to $[0, 2 \pi [$. 

Upon phase redefinitions for $\nu$ and $e_L$ fields and a unitary transformation for $e_R$ fields,  
one can go into the basis where the Lagrangian (\ref{lagr}) reads  
\beq
{\cal L} = -\frac{1}{2}~\nu^T ( {U^{(s)}_\nu}^* \hat m_\nu {\cal V_\nu}^{*2} {U^{(s)}_\nu}^\dagger )   \nu 
         - \bar e_R^{T} (\hat m_e {U^{L(s)}_e}^\dagger)  e_L 
         +~\frac{g}{\sqrt{2}}~ \bar e_L^T  \gamma^\lambda {\cal W}^* \nu ~W^-_\lambda +~ \mathrm{h.c.}~~~ , 
\label{tre}
\eeq
where ${\cal W}^*={\cal W}_e^*{\cal W}_\nu =$ diag$(e^{-i (w_1+w_2)}, e^{-i w_2},1)$.
The phases $w_2$ and $w_1$ can be chosen in $]-\pi,\pi] $ and represent the phase difference between 
the second and third generations of leptons, first and second respectively, in the basis (\ref{tre}), 
namely before shuffling the flavours by means of $U^{(s)}_\nu$ and $U^{L(s)}_e$ to go in the mass basis. 
They are a source for CP violation and,
in spite of the particular convention adopted here to define angles and 
phases\footnote{Needless to say, the standard parameterization for unitary matrices is as noble 
as other possible ones.}, 
it has to be recognized that they are univocally determined 
by the flavour symmetry and cannot be removed. 
Clearly $w_2$ and $w_1$ are not directly measurable, 
but nevertheless play a quite profound role for the MNS mixing matrix:
\bea
U_{MNS} & = & {U^{L(s)}_e}^\dagger {\cal W}^*  U^{(s)}_\nu {\cal V}_\nu \nn \\
        & = & \underbrace{R^T(\theta^e_{12}) \Gamma_{\delta^e} R^T(\theta^e_{13}) \Gamma_{\delta^e}^\dagger}_{S_e} 
       ~ \underbrace{R^T(\theta^e_{23}) {\cal W}^* R(\theta^\nu_{23}) }_{L} ~
   \underbrace{ \Gamma_{\delta^\nu} R(\theta^\nu_{13}) \Gamma_{\delta^\nu}^\dagger R(\theta^\nu_{12}){\cal V}_\nu }_{S_\nu}~~.
\label{twelve}
\eea
None of the 12 parameters dictated by the flavour symmetry and appearing in the r.h.s of eq. (\ref{twelve})
is actually measurable, because only 9 combinations of them are independent - see eq. (\ref{par}) -, 
among which 3 can be absorbed, $\alpha_{MNS}$ and ${\cal W}_{MNS}$.
Note that CP violation through $\delta$ can be generated in the limit where only $w_1$ and/or $w_2$ are present, 
as well as in the limit where there are just $\delta^e$ and/or $\delta^\nu$ 
(${\cal V}_\nu$ solely contributes to Majorana CP violating phases).  
Different flavour symmetries can thus predict the same leptonic
(and even hadronic) physics and there is no way to discriminate between them 
unless adopting theoretical criteria like, e.g., absence of tunings and stability of the results.
Adhering to such criteria, we now ask which sets of flavour symmetry parameters more robustly predict 
the atmospheric angle to be maximal.

It is natural to expect the leading role to be played by the core of the MNS mixing matrix, denoted by $L$ 
in eq. (\ref{twelve}):
\beq
L =\left( \matrix{  e^{-i( w_1+ w_2)} & 0 & 0 \cr
                    0 &  e^{- i w_2} c_e c_\nu + s_e s_\nu & e^{- i w_2}c_e s_\nu - s_e c_\nu   \cr
                   0 &  e^{- i w_2}s_e c_\nu -c_e s_\nu & e^{- i w_2} s_e s_\nu +c_e c_\nu } \right) 
\eeq
\\
where from now on $s_{e,\nu} = \sin \theta^{e,\nu}_{23}$, $c_{e,\nu} = \cos \theta^{e,\nu}_{23}$ for short.
If also $S_e$ in eq. (\ref{twelve}) had large mixings, bringing it at the right of $L$ would in general induce large
contributions to all three MNS mixings. 
Both the experimentally small $\theta_{13}$ and a potentially maximal $\theta_{23}$  
would then result from a subtle conspiracy between the many angles and phases 
involved. At the price of some tuning in the neutrino spectrum, 
this may happen in the case of tribimixing models \cite{A4}, which are often based on an $A_4$ flavour symmetry.
On the contrary, the bound on $\theta_{13}$ is naturally fulfilled if each mixing in $S_e$ and $R(\theta^\nu_{13})$ does. 
Denoting them with $\varphi = \sin \theta^e_{12}$, $\psi = \sin \theta^e_{13}$ and
$\xi = \sin \theta^\nu_{13}$, it turns out that they induce second order corrections in the 23 block of $L$,
%which are of second order in $\varphi, \psi, \xi$, 
hence smaller than $O(5\%)$ given the %present 
bound on $\theta_{13}$. 

Adopting the latter point of view, 
the atmospheric mixing angle can be identified with the one of $L$ up to corrections smaller than about $2^\circ$.
It turns out to depend on $w_2$ and, symmetrically, on $\theta^e_{23}$ and $\theta^\nu_{23}$:
\beq
\tan^2\theta_{23} = \frac{c_e^2 s_ \nu^2 + s_e^2 c_\nu^2 - 2 \cos w_2~ c_e s_e c_\nu s_\nu}
{s_e^2 s_ \nu^2 + c_e^2 c_\nu^2 + 2 \cos w_2 ~c_e s_e c_\nu s_\nu}~~~~~.
\label{tan2atm}
\eeq

\begin{figure}[!h]
\vskip .5 cm
\centerline{ \psfig{file=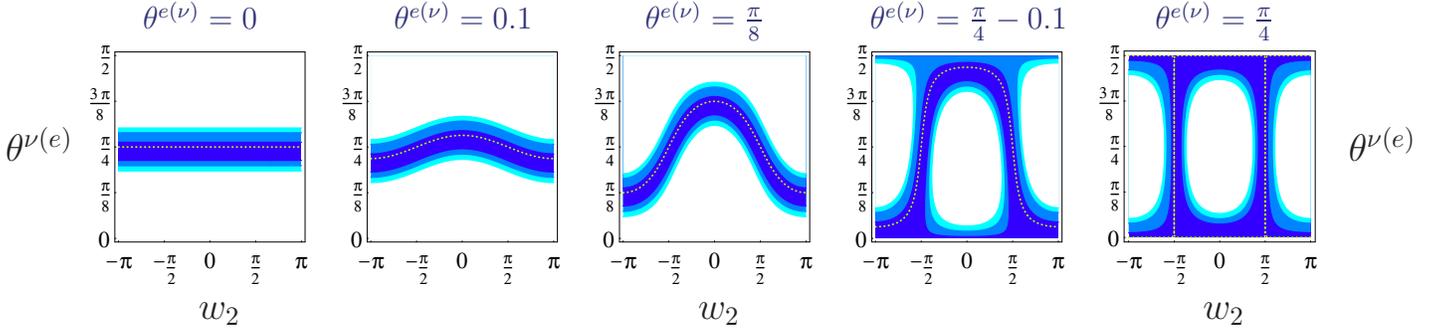,width=1.1 \textwidth} 
\Blue
\put(-470, 100){ $\theta^{e(\nu)}=0$}\put(-375, 100){ $\theta^{e(\nu)}=0.1$} \put(-280, 100){ $\theta^{e(\nu)}=\frac{\pi}{8}$}
\put(-195, 100){ $\theta^{e(\nu)}=\frac{\pi}{4} -0.1$}\put(-87, 100){ $\theta^{e(\nu)}=\frac{\pi}{4}$} 
\Black
\put(-518, 50){\large \bf $\theta^{\nu(e)}$} \put(-10, 50){\large \bf $\theta^{\nu(e)}$}
\put(-445, -10){\large \bf $w_2$}\put(-256, -10){\large \bf $w_2$}\put(-65, -10){\large \bf $w_2$}}
\caption{Region of the plane $\{ w_2,\theta^{\nu(e)}_{23} \}$ allowed at $1,2,3 \sigma$ by the experimental data on the 
atmospheric angle \cite{Fogli} for $\theta^{e(\nu)}_{23} =\{0, 0.1, \pi/8, \pi/4 -0.1, \pi/4\}$. 
The dotted curve is the surface where the atmospheric angle is maximal.
The plots also correspond to $\theta^{e(\nu)}_{23} = \pi / 2  -  \{0, 0.1, \pi/8, \pi/4 -0.1, \pi/4\}$ 
upon the substitution $\theta^{e(\nu)}\rightarrow \pi/2 - \theta^{e(\nu)}$.  }
\label{fig1}
\vskip .5 cm
\end{figure}

The crucial role played by the phase $w_2$ is manifest:
only in the case $w_2=^{~0}_{~\pi}$ one has the simple relation $\theta_{23}= |\theta^\nu_{23} \mp \theta^e_{23} |$.
A maximal atmospheric angle requires the following relation among three parameters to be fulfilled 
\beq
\cos w_2 = - \frac{1}{\tan(2 \theta^\nu_{23}) \tan(2 \theta^e_{23})}~~~.
\eeq
Notice that maximality is generically lost by slightly varying one of the parameters involved.
However, one realises from the above expressions that for some exceptional values of two parameters
the atmospheric angle turns out to be maximal 
independently from the value assumed by the third parameter: 
\vskip 0 cm 
{\bf i)} for $\theta^{e(\nu)}_{23} = \pi/4$ and $w_2 = \pm \pi/2$, independently of $\theta^{\nu(e)}_{23}$;
\vskip 0 cm
{\bf h)} for $\theta^{e(\nu)}_{23} = \pi/4$ and $\theta^{\nu(e)}_{23} = 0$ or $\pi/2$, independently of $w_2$. 
\vskip 0 cm
\noindent All this is graphically seen in fig. \ref{fig1} which, for different values of $\theta^{e(\nu)}_{23}$, 
shows the region of the plane $\{ w_2,\theta^{\nu(e)}_{23} \}$ allowed at $1,2,3 \sigma$ by
the experimental data on the atmospheric angle.

The above limits can be pictorially represented in terms of triangles.
In the case that $\theta^{e(\nu)}=\pi/4$, eq. (\ref{tan2atm}) can be rewritten under the form
\beq
\sqrt{2} \sin\theta_{23}=r=|  c_{\nu(e)}  - e^{-i w_2} s_{\nu(e)} |~~~~~
\label{tri}
\eeq
which is clearly reminiscent of a triangle, % of segments $r, c_{\nu(e)}, s_{\nu(e)}$, 
$w_2$ being the angle opposite to $r$. 
A maximal atmospheric angle requires $r=1$, as happens in the two cases below.
\begin{figure}[!h]
\vskip 0 cm
\centerline{ \psfig{file=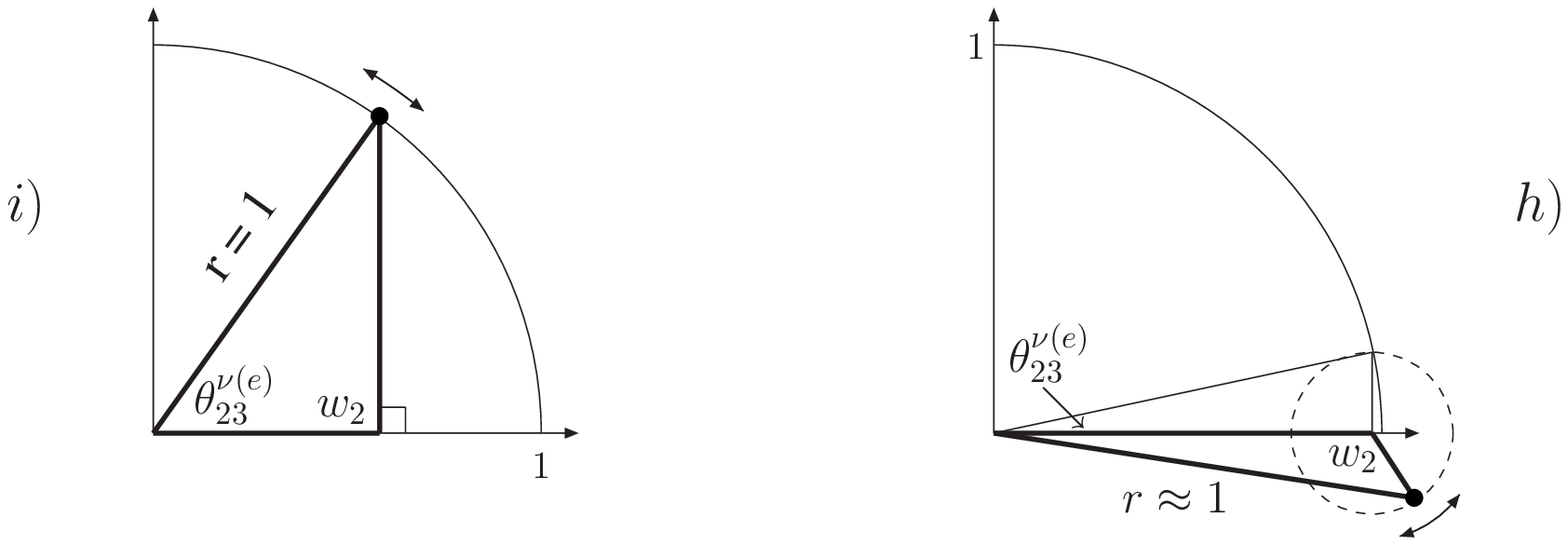,width=.8 \textwidth}  }
\label{trian}
\end{figure}

The possibility h) has been widely exploited in flavour model building.
The difficulty of this approach is not to predict a maximal $\theta^{e(\nu)}_{23}$,
but rather to manage in having a sufficiently small $\theta^{\nu(e)}_{23}$, say less than $2^\circ$.
For instance, a negligible $\theta^\nu_{23}$ is somewhat unnatural in the case of 
hierarchical neutrinos because the ratio between the corresponding eigenvalues is not so small: $m_2/m_3 \sim 1/6$.
Notice also that in seesaw models $\theta^{\nu}_{23}$ is an effective angle which depends on both the Dirac and 
Majorana Yukawa couplings. 

\begin{figure}[!ht]
\vskip .3 cm
\centerline{ \psfig{file=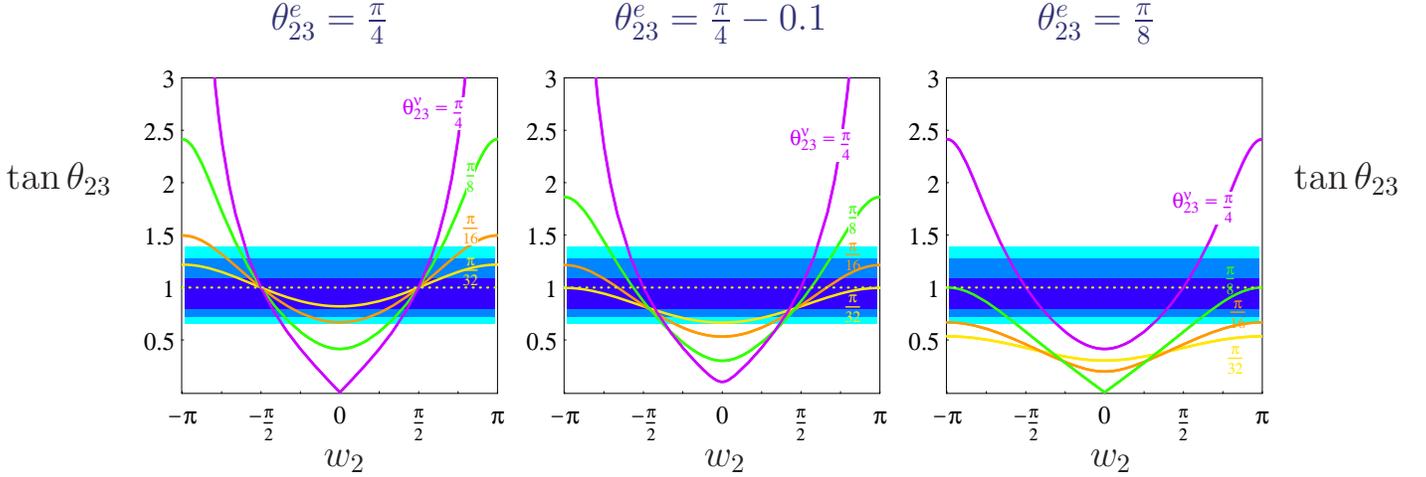,width=1 \textwidth}
\Blue
\put(-400, 160){ \large $\theta^e_{23}=\frac{\pi}{4}$}
\put(-270, 160){ \large $\theta^e_{23}=\frac{\pi}{4}-0.1$}
\put(-110, 160){ \large $\theta^e_{23}=\frac{\pi}{8}$}
\Black
\put(-500, 100){ \large $\tan\theta_{23}$} \put(-13, 100){ \large $\tan\theta_{23}$} 
\put(-380, -5){ \large $w_2$}\put(-236, -5){ \large $w_2$}\put(-90, -5){ \large $w_2$}}
\caption{ 
Values of $\tan\theta_{23}$ as a function of $w_2$ for $\theta^e_{23}=\pi/4,\pi/4-0.1,\pi/8$. 
Colored curves correspond, as marked, to $\theta^\nu_{23} =  \pi/32, \pi/16 $, $\pi /8, \pi/4 $. 
The experimental range at $1,2,3\sigma$ \cite{Fogli} is also shown.
The same plot holds for $e \leftrightarrow \nu$. }
\label{fig2}
\vskip .5 cm
\end{figure}

The possibility i) has not (to our knowledge) been singled out so far\footnote{We thank however the authors of ref. \cite{quater}
for pointing out some interesting analogies with one of their quaternion family symmetry models.}. 
It has the advantage that it does not require a huge hierarchy between $\theta^e_{23}$ and $\theta^\nu_{23}$. 
Indeed, in the case that
$g/\sqrt{2}~ (\bar \tau_L \gamma^\lambda \nu_\tau ~\pm ~i~ \bar \mu_L \gamma^\lambda  \nu_\mu)~ W^-_\lambda +$ h.c.
are the charged currect interactions of the second and third lepton families in the basis (\ref{tre}) - 
namely before applying $R(\theta^e_{23})$ and $R(\theta^\nu_{23})$ to go in the mass eigenstate basis -
the maximal phase difference $i$ shields a maximal $\theta^{e(\nu)}_{23}$ 
from any interference due to $\theta^{\nu(e)}_{23}$.
This remarkable fact can be visually seen in fig. \ref{fig2}, where we plot $\tan\theta_{23}$ 
as a function of $w_2$ for different values of $\theta^e_{23}$ and $\theta^{\nu}_{23}$.

It is worth to stress that 
a maximal CP violating phase difference in between fermion families 
is likely to be at work in the quark sector too, in particular for the Cabibbo angle \cite{Fri}
-  by the way, again the largest angle of the mixing matrix. Neglecting
the presumably very small $13$ and $23$ mixings, the charged currect interaction of the lighter families
in the basis analog of (\ref{tre}) is 
$g/\sqrt{2}~ (\bar c_L ~\gamma^\lambda s_L~+$ $ e^{-i w^q_1} ~\bar u_L ~\gamma^\lambda d_L)$ $~ W^+_\lambda +$ h.c.. 
Expanding at first order in $s_u=\sin \theta^u_{12}$ and $s_d=\sin \theta^d_{12}$, 
the analog of eq. (\ref{tan2atm}) reads 
\beq
\tan^2\theta_{C} =  s_u^2 + s_d^2  - 2 \cos w^q_1~  s_u  s_d + O(s^4)~~~~.
\eeq
In the plausible case that $s_d= \sqrt{m_d/m_s}$, $s_u= \sqrt{m_u/m_c}$, one has
\beq
|V_{us}|= | \sqrt{\frac{m_d}{m_s}}-e^{-i w^q_1} \sqrt{\frac{m_u}{m_c}} |
\eeq
and it is well known that experimental data strongly indicate $w^q_1= \pm \pi/2$.
Models of this sort have been studied where also $\alpha \approx w^q_1$ \cite{alfa}, 
so that the unitarity triangle turns out to be rectangular.

%%%%%%%%%%%%%%%%%%%%%%%%%%%%%%%%%%%%%%%%%%

\section{An Explicit Model with Maximal $w_2$}

We now sketch a supersymmetric $SO(3)$ flavour symmetry model 
which predicts a maximal atmospheric angle due to the presence of a maximal phase $w_2$.
The "flavon" chiral superfields and the lepton doublet $\ell$ are assigned to a triplet of $SO(3)$, 
while the lepton singlets $e^c,\mu^c,\tau^c$ and the Higgs doublets $h$ are $SO(3)$ singlets.
Along the lines of 
\cite{BHKR}\footnote{Actually, in ref. \cite{BHKR} models are discussed where $\theta_{23}=\pi/4$  
because $\theta^e_{23}=\pi/4$ and $\theta^\nu_{23}=0$.}
and in its spirit, interesting alignments for the flavon fields can be obtained.

Consider the superpotential:
\bea
W &=&  X_\chi ~(\chi^2 - M^2_\chi) + X_\varphi ~\varphi^2 + Y_{\chi \varphi}~ (\chi \varphi - M^2_{\chi \varphi})
~~~~~~~~~~~~~~~~~~~~~~~~~~~~(m^2_{\chi,\varphi,\phi}>0) \nn \\
  &+& X_\phi ~(\phi^2 - M^2_\phi) + Y_{\chi \phi}~ (\chi \phi -M^2_{\chi \phi})+ X_\xi ~\xi^2 + Y_{\chi \xi}~ \chi \xi 
~~~~~~~~~~~~~~~~~~( m^2_\xi<0) \nn \\
  &+&(\ell \chi)^2 h h + (\ell \phi)^2 h h + (\ell \xi)^2 h h  
  +  \tau^c (\ell \varphi) h + \mu^c (\ell \phi) h+ e^c (\ell \xi) h
\label{super}
\eea
where $X$'s and $Y$'s are $SO(3)$ singlet "driver" chiral superfields,  
$\chi$, $\varphi, \phi, \xi$ are "flavon" chiral superfields with positive soft mass squared except for $\xi$,
and dimensionful - eventually hierarchical - couplings are understood in the last line of eq. (\ref{super}). 
A thorough discussion of the discrete symmetries which could guarantee the previous couplings
and forbidding undesirable ones is beyond the spirit of the present discussion.
Minimization of the potential induces $SO(3)$ breaking by 
$<\chi>=(0,0,1)M_\chi$, $<\varphi>=(0,i,1)M^2_{\chi\varphi}/M_\chi$, 
$<\xi>=(i,1,0) M_\xi$ and $<\phi>=(0,\sin\alpha,\cos\alpha) M_\phi$, where $\cos \alpha= M_\chi M_\phi /M^2_{\chi\phi}$.
The following textures are then obtained
\beq
m_\nu \propto \left( \matrix{ -\lambda_\xi & i \lambda_\xi  & 0 \cr 
                      i \lambda_\xi & \lambda_\xi + \sin^2\alpha ~\lambda_\phi & \sin\alpha~\cos\alpha ~\lambda_\phi \cr 
                         0 & \sin\alpha ~\cos\alpha ~\lambda_\phi & \cos^2\alpha~ \lambda_\phi + \lambda_\chi } \right) ~~~~
m_e \propto  \left( \matrix{ i \epsilon_\xi & \epsilon_\xi & 0 \cr 0 &  \sin\alpha~ \epsilon_\phi & \cos\alpha~ \epsilon_\phi \cr 
                0 & i \epsilon_\varphi & \epsilon_\varphi } \right)
\eeq
which, as we now turn to discuss, can easily reproduce the experimental data. 

The spectrum of $m_e$ depends negligibly on $\alpha$ and 
is accomodated for $\epsilon_\xi:\epsilon_\phi:\epsilon_\varphi= \sqrt{2} m_e : 2 m_\mu : m_\tau$.
As for $m_\nu$, a hierarchical spectrum follows from taking 
$\cos\alpha=0.8$ and $\lambda_\xi:\lambda_\phi:\lambda_\chi=0.08:0.2:1$.
The latter values imply $w_2=-\pi/2$, $w_1=\pi$ and, for the charged lepton sector, 
$\theta^e_{23}=\pi/4+O( m^2_\mu/m^2_\tau)$, $\theta^e_{12}=\theta^e_{13}=\delta^e=0$, 
while for the neutrino sector
$\theta^\nu_{13}=0.4^\circ$, $\theta^\nu_{12}=34^\circ$, $\theta^\nu_{23}=6^\circ$, $\delta_\nu=v_2^\nu=0$, $v_1^\nu=\pi$.
Combining the latter 12 parameters to obtain the MNS mixing matrix - see eq. (\ref{twelve}) -, 
it turns out that, due to the maximal $w_2$, the atmospheric angle is also maximal, 
$\theta_{23}=\pi/4+O( m^2_\mu/m^2_\tau)$. In addition, $\theta_{12}=\theta^\nu_{12}$, $\theta_{13}=\theta^\nu_{13}$,
Majorana phases vanish but $\delta=\pi/2$. 
Note that such maximal CP violation in weak charged currents through $\delta$ has to be completely ascribed 
to the maximality of $w_2$.

%%%%%%%%%%%%%%%%%%%%%%%%%%%%%%%%%%%%%%%%%%%%%%%%%%%%%%%%%%%%%

\section{Some General Relations and Limiting Cases}

Here we discuss the general relations between the parameters in the basis (\ref{tre}) and the 
measurable quantities $|U_{e3}|$, $\theta_{12}$ and the MNS phase $\delta$,
under the assumptions that the bound on $\theta_{13}$ is naturally fulfilled 
because so do $\varphi = \sin \theta^e_{12}$, $\psi = \sin \theta^e_{13}$ and $\xi = \sin \theta^\nu_{13}$. 
This allows $S_e$ to commute with $L$  
and the dependence on the mechanism responsible for a maximal atmospheric angle 
is encoded in $w_2,\theta^e_{23},\theta^\nu_{23}$.  
Since $L_{22}=e^{- i w_2} L_{33}^*$, $L_{32}= - e^{- i w_2} L_{23}^*$ and a maximal atmospheric
angle implies $|L_{ij}|= 1/\sqrt{2}$ for $i,j=2,3$,
this dependence is equivalently expressed in terms of $w_2$, $\lambda_{23}=$Arg$(L_{23})$, $\lambda_{33}=$Arg$(L_{33})$.
We collect in table 1 the values of $\lambda_{23}, \lambda_{33}$ for the cases h) and i) discussed previously. 

\begin{table}[!h]
\begin{center}
\begin{tabular}{c||c|c|c|c} 
& $\theta^{e}_{23} = \pi/4$ & $\theta^{\nu}_{23} = \pi/4$ & $\theta^{e}_{23} = \pi/4$ & $\theta^{\nu}_{23} = \pi/4$ \\
& $\theta^{\nu}_{23} = 0 ~(\pi/2)$   &   $\theta^{e}_{23} = 0~(\pi/2)$     &       $w_2 = \pm \pi/2$   & $w_2 = \pm \pi/2$ \\ \hline
$\lambda_{23}$ & $\pi~ (-w_2)$ &  $- w_2~(\pi)$ & $\pm  \theta^\nu_{23} + \pi$ & $\mp  (\theta^e_{23}+\pi/2)$\\
$\lambda_{33}$ & $0 ~(-w_2)$ &  $0~(-w_2)$ & $\mp  \theta^\nu_{23}$ & $\mp  \theta^e_{23}$ \\
\end{tabular} \\ \vspace{.3 cm}
Table 1 
\end{center}
\end{table}

Introducing the quantities
\beq
v_\varphi= \frac{\varphi}{\sqrt{2}}~ e^{i (w_1-\lambda_{33})} ~,~~~
v_\psi=\frac{\psi}{\sqrt{2}}~ e^{i (w_1 -\lambda_{23}-\delta^e)} ~,~~~
v_\xi = \xi ~e^{-i (w_2+\lambda_{23}+\lambda_{33}+\delta^\nu)}
\eeq
one has~ $\theta_{23}= \pi/4 + O(v^2)$ with $v^2\sim 5\%$, together with the general formulas
\bea
\theta_{12} &=& \theta^\nu_{12} - \mathrm{Re}(v_\varphi - v_\psi) +O(v^2)  \nn\\ 
U_{e3} &=&  v_\varphi + v_\psi - v_\xi   + O(v^3) \\ 
\delta & = &  \pi - \mathrm{Arg}U_{e3} + O(v\sin(\mathrm{Arg}U_{e3})) ~~.\nn
\eea 
The above expressions allow to complete the phenomenological study of our framework
and generalise previous studies 
that assumed bimixing \cite{devdabimax} or tri-bimixing \cite{devdabitrimax} for $U_{\nu}$.
As already stressed, a measure of these independent observable quantities cannot reveal from which mechanism they come from,
nor whether the $v$'s interfere in originating them. 
In order to find some potential correlations, additional hypothesis have to be introduced.

The model of the previous section corresponds to the limit $\xi \gg \varphi,\psi$, in which case 
the above expressions simplify to
\beq
|U_{e3}|\approx \xi ~~,~~~~\delta \approx w_2+\lambda_{23}+\lambda_{33}+\delta^\nu ~~,
~~~~ \theta_{12} \approx \theta^\nu_{12} ~. 
\eeq 
Notice that there are no correlations between $|U_{e3}|$, $\theta_{12}$ and $\delta$. 
The latter does not depend on $w_1$ and is rather related to the mechanism at work for the atmospheric angle, 
even though it cannot reveal which one is actually at work. As can be seen from table 1,
in the case i) with $\theta^{e}_{23} = \pi/4$, $\delta=w_2+\pi+\delta^\nu$.
The model of the previous section had $\delta^\nu=0$ and $w_2=-\pi/2$, which explicitely shows 
that a maximal $w_2$ was the source of the maximal CP violation in $\delta$.

Interesting correlations emerge for $\psi\gg \varphi,\xi$ in which case  
\beq
|U_{e3}| \approx \frac{\psi}{\sqrt{2}} ~~,~~~~~\delta  \approx  \pi +\lambda_{23} +\delta^e - w_1 ~~,~~~~
\theta_{12}  \approx \theta^\nu_{12}  -   |U_{e3}| \cos \delta~, 
\eeq
and for $\varphi\gg \psi,\xi$ in which case  
\beq
|U_{e3}| \approx \frac{\varphi}{\sqrt{2}} ~~,~~~~~\delta  \approx  \pi +\lambda_{33} - w_1 ~~,~~~~
\theta_{12}  \approx \theta^\nu_{12}  +   |U_{e3}| \cos \delta ~.
\eeq
Notice that these situations are phenomenologically equivalent provided $\delta \leftrightarrow \delta +\pi$.
Both $\delta$ and $\theta_{12}$ depend on the mechanism at work for the atmospheric angle.
For instance, for $\varphi\gg \psi,\xi$ and with $\theta^{e(\nu)}_{23} = \pi/4$, $w_2 = \pm \pi/2$, 
it turns out that $\delta$ depends on $w_1$ and the 23-angle whose magnitude is irrelevant for the atmospheric angle: 
$\delta \approx  \pi - w_1 \mp \theta^{\nu(e)}_{23}$.

\begin{figure}[!h]
\vskip 0. cm
\centerline{ \psfig{file=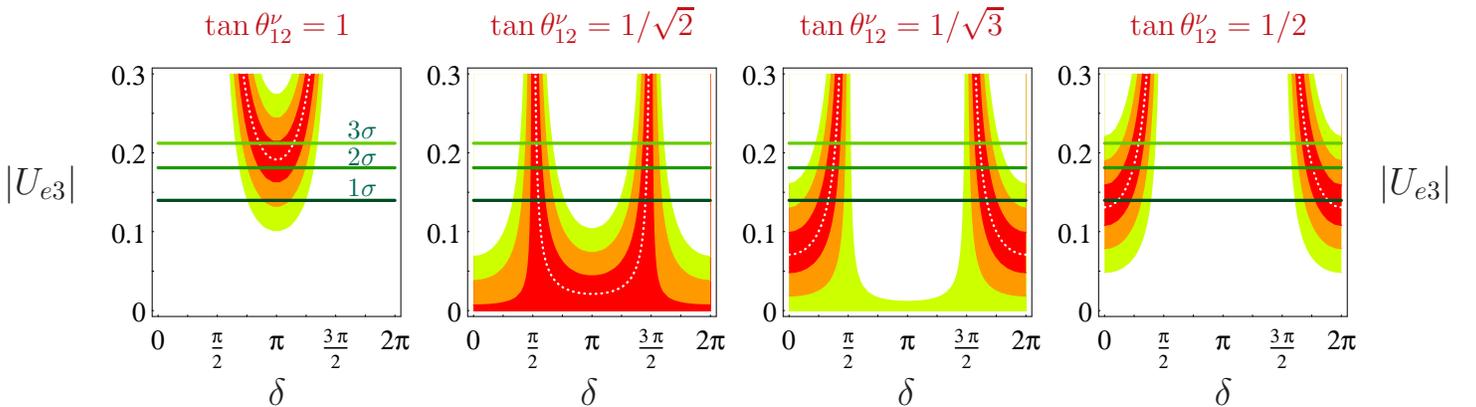,width=1.1 \textwidth}
\BrightRed
\put(-458, 130){ $\tan\theta^{\nu}_{12}=1$} \put(-350, 130){ $\tan\theta^{\nu}_{12}=1/\sqrt{2}$} 
\put(-233, 130){ $\tan\theta^{\nu}_{12}=1/\sqrt{3}$} \put(-108, 130){ $\tan\theta^{\nu}_{12}=1/2$}
\Green
%\put(-460, 55){ \Large $\uparrow $} \put(-470, 37){ \large \bf QLC}
\put(-400, 91){\footnotesize $3\sigma$}\put(-400, 81){\footnotesize $2\sigma$}\put(-400, 69){\footnotesize$1\sigma$}
\Black
\put(-530, 70){\large \bf $|U_{e3}|$} \put(-10, 70){\large \bf $|U_{e3}|$}
\put(-430, -10){\large \bf $\delta$}\put(-312, -10){\large \bf $\delta$}
\put(-192, -10){\large \bf $\delta$}\put(-72, -10){\large \bf $\delta$}   }
\caption{Region of the $\{ \delta, |U_{e3}|\}$ plane allowed by the present range of $\theta_{12}$ at $1,2,3 \sigma$ 
\cite{Fogli}
for different values of $\tan\theta^\nu_{12}$. The dotted line corresponds to the best fit of $\theta_{12}$.
Also shown are the $1,2,3 \sigma$ bounds on $|U_{e3}|$.}
\label{fig3}
\vskip 0. cm
\end{figure}

In the following we focus on the possibility that $\sin \theta^e_{12}= \varphi$ dominates. 
This is a particularly interesting scenario because naturally compatible with a grandunified picture. 
The correlations are shown in fig. \ref{fig3}  
by plotting, for different values of $\theta^\nu_{12}$,
the region of the $\{ \delta, |U_{e3}|\}$ plane allowed by the present range of $\theta_{12}$ at $1,2,3 \sigma$.  
The case of a maximal $\theta^\nu_{12}$ is particularly interesting from the theory point of view. 
As shown by the plot, present data \cite{Fogli} suggest $\delta \approx \pi$ 
and $|U_{e3}|\approx 0.2$, dangerously close to its $3\sigma$ bound and interestingly close to the Cabibbo angle $\theta_C$.
Notice that  
the so-called "quark lepton complementarity" proposal $\theta_{12}=\pi/4- \theta_C$ \cite{qlc}
corresponds precisely to  $\delta = \pi$ and $|U_{e3}|= \theta_C$, i.e. exact CP and $\varphi= \sqrt{2} \theta_C$.  
Anyway, also $\varphi=\theta_C$, i.e. $U_{e3}$ close to its $1\sigma$ bound,
falls inside the $2\sigma$ window for $\theta_{12}$ provided $\delta = \pi$.
Remarkably enough, it turns out that a maximal $\delta$ strongly favours
$\tan\theta^{\nu}_{12} \approx 1/\sqrt{2}$, with a mild dependence on $|U_{e3}|$.
The possibility $\varphi=\theta_C/3$, particularly relevant for grandunified models, 
is thus well compatible with tribimixing and maximal CP violation.

%%%%%%%%%%%%%%%%%%%%%%%%%%%%%%%%%%%%%%%%%%%%%%%%%%%%%%%%%%%%%

\section{Conclusions and Outlook}

We pointed out that if CP and flavour are maximally violated by
second and third lepton families in the flavour symmetry basis,
a maximal atmospheric angle is automatically generated 
when the bound on $|U_{e3}|$ is fulfilled in a natural way.
This mechanism has two advantages with respect to the one usually exploited:
it is very suggestive of the quark sector and it does not require one between $\theta^\nu_{23}$ and $\theta^e_{23}$
to vanish, which could be difficult to achieve especially for seesaw models. 

We think that such a mechanism deserves more studies, 
both from the point of view of grandunified theories and flavour symmetries.

Under the assumption that the bound on $|U_{e3}|$ is naturally fulfilled,
we discussed the general relations between the parameters in the basis (\ref{tre}) and the 
measurable quantities $|U_{e3}|$, $\theta_{12}$ and the CP violating phase $\delta$,
clarifying in particular its relation with the phases among lepton families.
These general results have also been confronted with the preditions of a specific realisation 
of the above mechanism, a supersymmetric model based on a $SO(3)$ flavour symmetry where 
a maximal CP violating phase $\delta$ arose as a direct consequence of the maximal phase difference between second
and third lepton families.

%%%%%%%%%%%%%%%%%%%%%%%%%%%%%%%%%%%%%%%%%%%%%%%%%%%%%%%%%%%%%
\section*{Acknowledgements}

I thank G. Altarelli and C.A. Savoy for enlightening discussions.
Thanks also go to the Dept. of Physics of Rome1 and to the SPhT CEA-Saclay 
for hospitality during the completion of this work. 
This project is partially supported by the RTN 
European Program MRTN-CT-2004-503369.

%%%%%%%%%%%%%%  Bibliography   %%%%%%%%%%%%%%%%%%%%%%%%%%%%%%%

%%%%%%%%%%%%%%%%%%%%%%%%%%%%%%%%%%%%%%%%%%%%%%%%%%%%%%%%%%%

\end{document}